\documentclass[10pt,twocolumn]{article}
\usepackage{graphicx}

\begin{document}
\title{Giant magnetoresistance oscillations caused by cyclotron
resonance harmonics}
\author{S.~I.~Dorozhkin\\
Institute of Solid State Physics, Chernogolovka, Moscow
district,142432, Russia}

\date{\today}

\maketitle

\begin{abstract}
\noindent
For high-mobility two-dimensional electrons at a GaAs/AlGaAs
heterojunction, we have studied, both experimentally and
theoretically, the recently discovered giant magnetoresistance
oscillations with nearly zero resistance in the oscillation minima
which appear under microwave radiation. We have proposed a
model based on nonequilibrium occupation of Landau levels caused by
radiation which describes the oscillation picture.
\end{abstract}


Recent observations, in high-quality two-dimensional electron
systems, of microwave-stimulated giant magnetoresistance
oscillations (MSGMO)~\cite {zudov1,ye} with positions
corresponding to subharmonics of the cyclotron
resonance~\cite{comm} and especially the discover of
zero-resistance states in the MSGMO minima~\cite{mani,zudov2}
attract a great attention to this spectacular
phenomenon~\cite{phil,durst,andreev,ander,shi,volkov,koul,shikin,mikh}.
Observation of zero-resistance states was a reason to
assume~\cite{mani,zudov2} their collective origin, like
photon-stimulated superconductivity~\cite{mani}.  Alternative
approach to explanation of zero-resistance states is based on the
peculiarities of electron motion in crossed electric and magnetic
fields, when electron drift along electric field can occur only as
a result of scattering events. In recent preprint~\cite{durst}, it
has been demonstrated that MSGMO with negative magnetoresistance
in minima can result from transitions between broadened Landau
levels caused by photon absorption and accompanied by elastic
scattering on short-range scatterers. Similar effect was shown
rather long ago ~\cite{ryzhii} to give a sequence of photocurrent
peaks of different signs for unbroadened Landau levels and
nonlinear conditions with respect to electric field. For bulk
semiconductors and quantising magnetic fields, photocurrent
oscillations with negative conductivity in minima were predicted
in Ref.~\cite{elesin} for $\delta$ -type photoelectron energy
distribution function. A link between states with negative
dissipative conductivity and the zero-resistance states was
proposed in Ref~\cite{andreev} (see also Ref.~\cite{volkov}). It
implies that negative dissipative resistivity gives rise to
instability in a system which finally breaks its symmetry and
produces inhomogeneous states with nearly zero average resistance.
This result allows to associate theoretical states with negative
dissipative resistivity and the experimental zero-resistance
states. The inhomogeneous states are characterized by high local
current density even at zero net current through a sample. One
more approach to explanation of MSGMO based on edge magnetoplasma
instability was considered in preprint~\cite{mikh}.

In this paper, we reproduce previous experimental observations~\cite
{zudov1,ye,mani,zudov2} of MSGMO with some additional data and compare them
with our calculations based on results of self-consistent Born approximation
(SCBA) which are capable of explaining the main features of MSGMO in terms of
nonequilibrium occupation of broadened Landau levels under microwave
radiation. Additionally, our model predicts appearance of the second harmonic
in the Shubnikov-de Haas oscillations under appropriate choice of microwave
frequency and sample parameters.

We have measured a Hall bar sample with conducting channel of a
L-shape. Channel width was equal to 0.2 mm, the distances between
neighboring potential probes were either 0.4 or 0.6 mm.
Magnetoresistance per square $R_{\rm xx}$ and Hall resistance
$R_{\rm xy}$ have been measured by the standard technique
exploiting the low-frequency AC current excitation and phase
sensitive detection of a voltage between potential probes with the
use of a Lock-in amplifier. We used the frequency 9.2 Hz and
amplitude of the current 1 $\mu A$, well within ohmic regime.
Results for $R_{\rm xx}$ and $R_{\rm xy}$ presented in this paper
do not depend on pairs of potential probes used for measurements.
The sample has been produced from a GaAs/AlGaAs wafer of the
standard architecture containing a two-dimensional electron system
at a single remotely doped GaAs/AlGaAs heterojunction with the
spacer width about 55 nm. The most pronounced MSGMO were observed
after illumination of the sample until saturation of electron
density at about $n_{{\rm s}}=3\times 10^{11}{\rm cm}^{-2}$. The
corresponding mobility of electrons was $\mu=7\times 10^{6}{\rm
cm}^2/{\rm V\,s}$. The sample was placed in the close ended
rectangular waveguide of the WR-28 type with cross-section sizes
$16\times 8\mbox{mm}^2$ which entered a $^3{\rm He}$ refrigerator.
The two arms of the L-shaped sample were parallel to the long and
the short sides of the rectangular. The microwave radiation in the
frequency range 10 -170 GHz was produced by a set of oscillators.
The microwave power reached the low temperature end of the
waveguide was estimated to be always below 2 mW.  At frequencies
greater than 50 GHz, transmission coefficient between oscillator
output and the low-temperature part of the waveguide was rather
low falling down to values of the order of 0.01.

\begin{figure}[h]
\includegraphics[width=0.45\textwidth,clip]{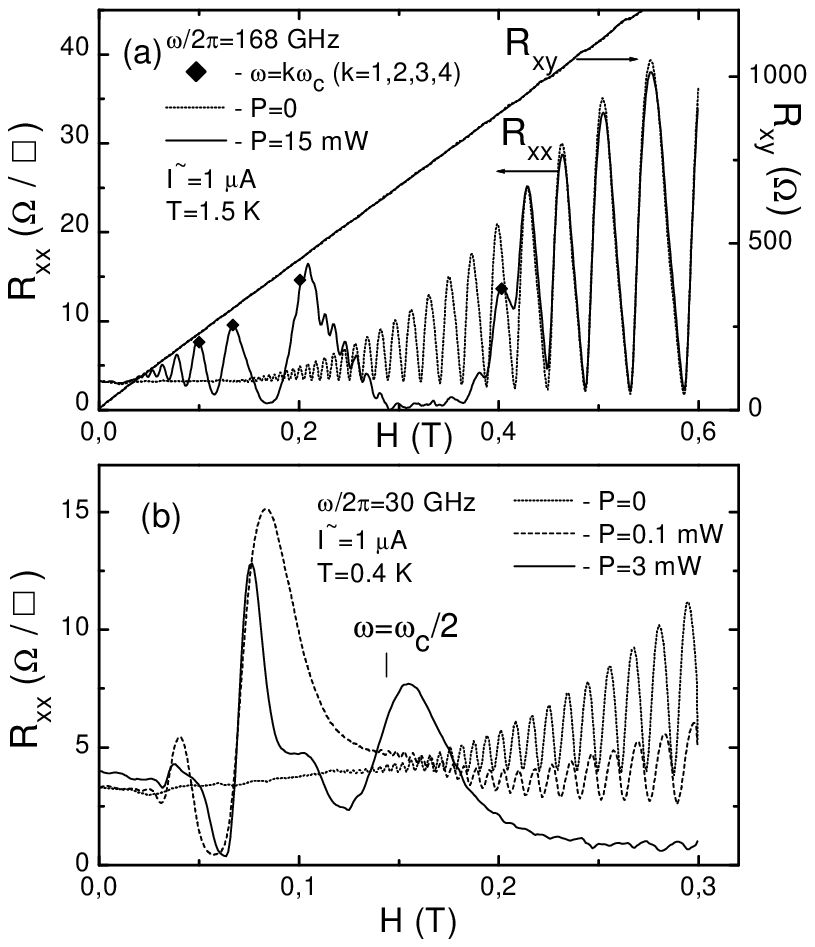}
\caption{Magnetoresistance $R_{\rm xx}$, Hall resistance $R_{\rm
xy}$ measured with excitation current $I^{\sim}=1\,{\rm
\protect\mu A}$ versus magnetic field $H$ in the absence of
microwave radiation (dotted lines) and under microwave radiation
(dashed and solid lines). Data in Figs 1a and 1b have been
measured at frequencies 168 GHz and 30 GHz, respectively. Power
$P$ shown in Figs. corresponds to the oscillator output. Positions
of magnetic fields corresponding to subharmonics of the cyclotron
resonance are marked in Fig. 1a by diamonds. $n_{\rm s}=2.84
\times 10^{11} {\rm cm}^{-2}$.}
\end{figure}

Typical experimental data are shown in Fig.1. In the absence of
microwave radiation, magnetoresistance $R _{\rm xx}$ demonstrates
at $H\geq 1.5$~T standard Shubnikov-de Haas oscillations periodic
in the inverse magnetic field with the period determined by areal
density of electrons $n_{\rm s}$ in a two-dimensional system.
Corresponding oscillations in $R _{\rm xy}$ become visible only at
magnetic fields higher than 0.4 T. Microwave radiation suppresses
Shubnikov-de Haas oscillations at low magnetic fields and gives
rise to new oscillations (microwave stimulated giant
magnetoresistance oscillations) also periodic in the inverse
magnetic field with the period determined by microwave frequency
(compare Fig.1a and Fig.  1b). Positions of MSGMO follow those of
subharmonics of the cyclotron resonance $\omega =\omega _{\rm
c}^{\rm (k)}\equiv k\,(eH^{\rm (k)}/m^{*}c)$. Here
$m^{*}=0.067m_{\rm e}$ is the effective mass of electrons in GaAs.
The main minima and maxima of MSGMO are shifted to different sides
from the corresponding subharmonic. Additional weaker oscillation
arises at comparatively low-frequency and high-power radiation at
$\omega <\omega_{\rm c}$ and can be associated with the cyclotron
resonance harmonic $\omega =\omega _{\rm c}/2$ (see Fig.1b). In
the main MSGMO minima, the magnetoresistance can become rather
close to zero (see also Refs [3,4]).  MSGMO peaks have
characteristic asymmetric triangular-like form with a steep drop
at low-magnetic-field side. At the same time, the microwave
radiation has practically no effect on the Hall resistance (in
Fig.1a the solid $R_{\rm xy}$ curve measured in the presence of
the microwaves is practically indistinguishable from the ''dark''
dotted curve).

Our calculations of magnetoconductivity tensor components
$\sigma_{\rm xx}$
and $\sigma_{\rm xy}$ are based on formulas obtained within
self-consistent Born approximation in the absence of Landau level mixing
(see review~\cite{ando} and Refs. therein):

\begin{equation}
D(\epsilon)=\sum_{n=1}^\infty \frac{2N_0}{\pi\Gamma_{{\rm n}}} \left[1-\left(
\frac{\epsilon-\epsilon_{{\rm n}}}{\Gamma_{{\rm n}}}\right)^2\right]^{1/2}
\equiv \sum_{i=1}^\infty \frac{2N_0}{\pi\Gamma_{{\rm n}}}Z_{{\rm n}
}^{1/2}(\epsilon)
\end{equation}

\begin{equation}
\sigma_{{\rm xx}}= \frac{e^2}{\pi^2 \hbar} \sum_{n=0}^{\infty} \left(\frac{%
\Gamma^{{\rm xx}}_{{\rm n}}}{\Gamma_{{\rm n}}}\right)^2 \int_{\epsilon_{{\rm %
n}}-\Gamma_{{\rm n}}}^{\epsilon_{{\rm n}}+\Gamma_{{\rm n}}} \left(-\frac{df}{%
d\epsilon}\right) Z_{{\rm n}}(\epsilon) d\epsilon
\end{equation}

\begin{equation}
\sigma_{{\rm xy}}= -\frac{n_{{\rm s}}ec}{H}+ \frac{e^2}{\pi^2 \hbar}
\sum_{n=0}^{\infty} \frac{\left(\Gamma^{{\rm xy}}_{{\rm n}}\right)^4}{%
\Gamma_{{\rm n}}^3\hbar\omega_{{\rm c}}} \int_{\epsilon_{{\rm n}}-\Gamma_{%
{\rm n}}}^{\epsilon_{{\rm n}}+\Gamma_{{\rm n}}} \left(-\frac{df}{d\epsilon}%
\right) Z_{{\rm n}}^{3/2}(\epsilon) d\epsilon
\end{equation}

\noindent Here $D(\epsilon)$ is the density of states, $\epsilon_{{\rm n}%
}=\hbar\omega_{{\rm c}}(n+1/2)$ is the energy of the n-th spin-degenerate
Landau level with the width $\Gamma_{{\rm n}}$ and the total number of
states on the level $N_0=2eH/hc$. Contribution of this level to $\sigma_{%
{\rm xx}}$ and $\sigma_{{\rm xy}}$ are characterized by parameters $\Gamma^{%
{\rm xx}}_{{\rm n}}$ and $\Gamma^{{\rm xy}}_{{\rm n}}$, respectively. Our
modification of Eqs. of Ref.~\cite{ando} is the use of nonequilibrium
distribution function $f(\epsilon)$ which would be formed, at zero
temperature, as a result of direct one-photon transitions (induced and
spontaneous), i.e., transitions accompanied by the energy change by
$\hbar\omega$ and the nearly zero momentum variation equal to the photon
momentum. We neglect all other excitation and relaxation processes.
Such function can arise if the life time of a photo-
excited electron is the shortest time in the problem. This condition is
normally justified if the energy of this electron is less than the energy
of optical phonon (see, for example, Ref~\cite{habb}), which is well
fulfilled in our experiment. As can be shown, the distribution
function obtained under these conditions is appropriate
for calculations of the magnetoconductivity
in accordance with Eqs.(2,3).

At zero temperature, it is easy to write down condition of the steady state
which relates values of the distribution function $f(\epsilon)$  at energies
differing by $\hbar\omega$:

\begin{equation}
f(\epsilon)= \frac{\lambda f(\epsilon-\hbar\omega)}{\lambda+1-f(\epsilon-%
\hbar\omega)}
\end{equation}

\noindent Here parameter $\lambda $ characterizes microwave intensity. This
relation is applicable at non-zero densities of states $D(\epsilon )$ and $%
D(\epsilon -\hbar \omega )$. If $D(\epsilon )=0$ or $D(\epsilon -\hbar
\omega )=0$ we set $f(\epsilon )=f_{0}(\epsilon )$, where $f_{0}$ is the
Fermi distribution function at $T=0$. Eq. (4) and condition
$\int\limits_{-\infty }^{+\infty }f(\epsilon )D(\epsilon )d\epsilon =n_{s}$
define nonequilibrium distribution function which was used for calculations
of the conductivity. Fig.2 demonstrates that,
in some ranges of energy, photon-stimulated interlevel
transitions can give rise to inverted population of electron states
($df/d\epsilon >0$) leading to negative contributions to the conductivity
$\sigma _{xx}$.  The inversion is possible only if $\omega >\omega _{{\rm
c}}$. The inverted occupation shown in the right panel of Fig.2 is
obviously independent of the position of the Fermi energy on the lower level,
i.e., of the filling factor of Landau levels. Our computation shows (see
Fig.3) that appearance of energy regions with inverted population of
broadened Landau levels can lead to the negative sign of $\sigma _{{\rm xx}}$
and, consequently, to the negative magnetoresistance $R _{{\rm xx}}=\sigma
_{{\rm xx}}/(\sigma _{{\rm xx}}^{2}+\sigma _{{\rm xy}}^{2})$. We have
considered two limiting cases of the short- and long-range scatterers when
analytical formulas are available for dependencies of parameters $\Gamma _{%
{\rm n}}$, $\Gamma _{{\rm n}}^{{\rm xx}}$, and $\Gamma _{{\rm n}}^{{\rm xy}}$
on $n$ and magnetic field~\cite{ando}. For the short-range (long-range)
scatterers there is only one (two) independent parameter(s). Fig. 3 shows
results of our calculations.  To get data corresponding to the absence of
radiation we have set parameter $\lambda $ to very low value $\lambda
=1\times 10^{-10}$. The results obtained under radiation provide MSGMO with
both, form of the oscillations and their positions, in reasonable agreement
with the experiment (namely, the minimum (maximum) associated with particular
subharmonic lies at $\omega >k\omega _{\rm c}^{\rm(k)}$ ($\omega<k\omega_{\rm
c}^{\rm(k)}$)). The main difference from the experiment is in the fact that
calculated magnetoresistance in the MSGMO minima is negative.  Elimination of
this discrepancy lies beyond our model applicable for macroscopically
homogeneous systems and could be referred to results of Ref.~\cite{andreev}.
Additionally, there are at least two more mechanisms leading to suppression
of the negative conductivity which are finite temperature and all kinds of
relaxation processes. These arguments show that the calculated regions of
negative magnetoresistance can be associated with the experimental minima of
MSGMO.  For $\omega =k\omega_{\rm c}^{\rm (k)}$ our model predicts absence of
the photoresponse in magnetoresistance. But this is valid only for the two
limiting cases of short-range and long-range scatterers, when the width of a
Landau level $\Gamma_{\rm n}$ is independent of a level number
$n$~\cite{ando}. For the intermediate-range scatterers, points where $R_{\rm
xx}|_{\rm P=0}=R_{\rm xx}|_{\rm P\neq 0}$ should depend on the radiation
power $P$ and be shifted from the positions of the subharmonics, which is
consistent with our experimental data where these points appear always at
lower magnetic fields than the corresponding subharmonics.  Additional shift
of these points can result from relaxation processes.

\begin{figure}[h]
\includegraphics[width=0.45\textwidth,clip]{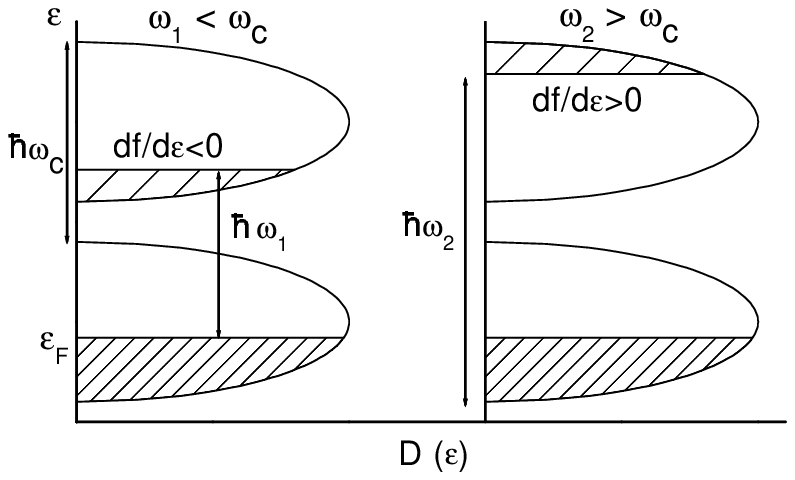}
\caption{ Schematic of the electron redistributions between two
neighboring broadened Landau levels which are caused by microwave
radiation of two frequencies $\omega_1<\omega_{\rm c}$ (left
panel) and $\omega_2>\omega_{\rm c}$ (right panel). Partly
occupied states are shaded. The lowest level contains the Fermi
energy $\epsilon_{\rm F}$ at zero temperature. Assumptions
concerning relaxation processes are discussed in the text.}
\end{figure}

\begin{figure}[h]
\includegraphics[width=0.45\textwidth,clip]{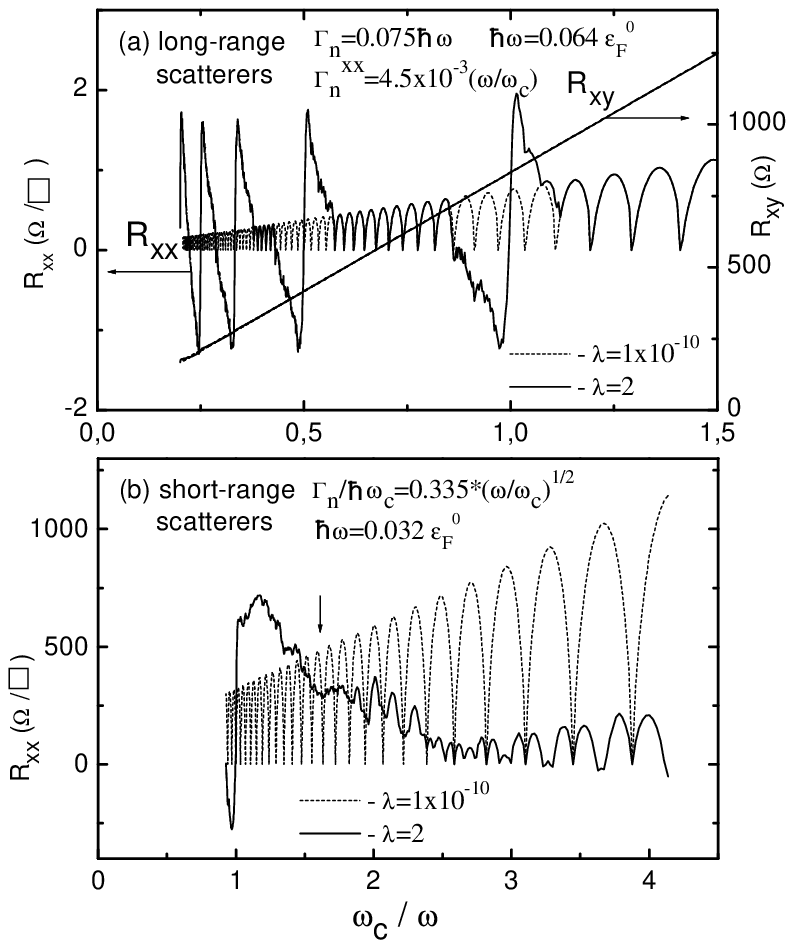}
\caption{Calculated magnetoresistance $R_{\rm xx}$ and Hall
resistance $R_{\rm xy}$ versus ratio $\omega _{\rm c}/\omega $
proportional to magnetic field for two very different intensities
of microwave radiation, characterizing by parameter $\lambda$
(dotted and solid lines). The solid and dotted $R_{\rm xy}$ lines
overlap. Figs. 3a and 3b correspond to two limiting cases of the
scatterer range, different values of the microwave frequency, and
different values of Landau level width (corresponding parameters
are shown in the Figure).}
\end{figure}

It is necessary to discuss a role of different unknown parameters
which enter our model.  Absolute values of $\Gamma_{\rm n}$,
$\Gamma_{\rm n}^{\rm xx}$, $\Gamma_{\rm n}^{\rm xy}$, and
$\lambda$ affect only amplitude and detailed form of the
Shubnikov- de Haas oscillations and MSGMO. Neither of these
parameters influences positions of the oscillations. Explanation
of the experimentally established absence of the photoresponse
in the Hall resistance is closely related to absolute values of parameters $%
\Gamma _{{\rm n}}$, $\Gamma _{{\rm n}}^{{\rm xx}}$, and $\Gamma _{{\rm n}}^{%
{\rm xy}}$. Within SCBA this result appears rather naturally for the case of
long-range scatterers (in Fig.3a the solid and dotted $R _{{\rm xy}}$
lines overlap) and needs very narrow Landau levels in the other limiting
case. A problem arises with absolute values of the magnetoresistance even in
the absence of microwaves (see also results of Ref.~\cite{durst} obtained
within SCBA for short-range scatterers). In comparison with experiment,
SCBA gives either too high or too low values of magnetoresistance for two
limiting cases of short-range and long-range scatterers, respectively. But it
seems to be possible to get reasonable absolute values of the
magnetoresistance together with the absence of the photoresponse in the Hall
resistance for intermediate-range scatterers. Comparison of the lower limit
for the range of potential fluctuations, given by the spacer width, with the
cyclotron radius in magnetic fields involved shows that in our experimental
conditions intermediate-range scatterers are of great importance.

Important aspect of the experimental results is existence of MSGMO in very
weak magnetic fields where Shubnikov-de Haas oscillations are not observed.
Obviously, appearance of MSGMO proves existence of the Landau quantization
in corresponding magnetic fields. We explain absence of the Shubnikov - de
Haas oscillations in these fields by inhomogeneous broadening of the
oscillations picture caused by long-range carrier density fluctuations in a
sample. MSGMO are much less affected by this broadening because of larger
period of these oscillations measured in filling factors of Landau levels.
Inhomogeneous broadening is not included in our model and Shubnikov - de Haas
oscillations and MSGMO coexist at the same magnetic fields.

It is interesting to note that, in addition to SMGMO, results of
our calculations describe some additional features of the
experimental data. In the case of comparatively narrow Landau
levels, the Shubnikov -  de Haas oscillations are only slightly
modified by microwave radiation at $\omega < \omega _{{\rm c}}$
and in-between $k=1$ minimum and $k=2$ maximum of MSGMO (compare
Figs 1a and 3a). At rather wide Landau levels, experimentally
realized at weaker magnetic field, our model describes strong
suppression of the magnetoresistance at $\omega \ll \omega _{{\rm
c}}$ in very good agreement with our observations (compare Figs.
1b and 3b) and
leads to appearance of additional minima related to the $\omega =\omega _{%
{\rm c}}/2$ harmonic (shown by arrow in Fig.3b). Note that the latter effect
appears in our model without two-photon processes. Our
model predicts that, for appropriate choice of Landau level width, microwave
frequency and power, great second harmonic can appear in the Shubnikov-de
Haas oscillation picture (in Fig 3b it occurs at
$\omega _{{\rm c}}/\omega>2.5$).

In summary, we have shown that experimentally measured photoresponse of
two-dimensional electrons on microwave radiation, including MSGMO, is
consistent with our theory considering conventional magnetotransport effects
under conditions of nonequilibrium occupation of electronic states.

The author is very thankful to Professor K. von Klitzing and Dr. J.H.
Smet for kind opportunity to carry out the experimental part
of this research in the Max-Planck-Insitut f\"{u}r
Festk\"{o}rperforschung (Stuttgart, Germany) and to Dr. V.Umansky
for providing the author with nice GaAs/AlGaAs material. The author
gratefully acknowledges valuable discussion with S.V.Iordansky and
partial support of this work by INTAS and RFBR.


\begin{thebibliography}{99}
\bibitem{comm} In this paper we use definition of the cyclotron
resonance subharmonics as $\omega=k\omega_{\rm c}^{\rm(k)}$ in accordance
with that in review~\cite{ando}. Here $\omega/2\pi$ - is the microwave
frequency and $\omega_{\rm c}=eH/m^*c$ is the cyclotron frequency in magnetic
field $H$ for carriers with effective mass $m^*$.
\end{thebibliography}
\end{document}